\documentclass[preprint,showpacs,preprintnumbers,nofootinbib,prd]{revtex4}
\usepackage{amsmath,amssymb}
\usepackage[dvips]{graphicx}
\usepackage{dcolumn}
\usepackage{bm}

\begin{document}

\title{Muon anomalous magnetic moment \\ due to the brane-stretching effect}

\author{Konosuke Sawa}
\affiliation{Department of Physics, Tokyo University of Science, 1-3 Kagurazaka, Shinjyuku-ku, Tokyo 162-0861, Japan}%

\begin{abstract}
We investigate the contribution of extra dimensions to the muon anomalous magnetic moment by using an ADD-type $6$-dimensional model. This approach analyzes the extent of the influence of classical brane fluctuations on the magnetic moment. When we consider that the brane fluctuations are static in time, they add new potential terms to the Schr{\"o}dinger equation through the induced vierbein. This paper shows that the brane fluctuation is responsible for the brane-stretching effect. This effect would be capable of reproducing the appropriate order for recent Brookhaven National Laboratory measurements of the muon (g-2) deviation. 
\end{abstract}


\maketitle

\section{Introduction}\label{section 1}
The BNL E821 group recently reported the precision measurement of the muon anomalous magnetic moment. Based on their result, a new world average recorded $a^{(\textrm{exp})}_{\mu}=11659208(6)\times 10^{-10} (\pm 0.7 \textrm{ppm})$ \cite{Brown:2000sj}, whereas H{\"o}cker et al. obtained the Standard Model (SM) prediction $a^{(\textrm{SM})}_{\mu}=11659182(6)\times 10^{-10} (\pm 0.7 \textrm{ppm})$ \cite{Hocker:2004xc}. The difference in values, $\Delta a_{\mu}\equiv a^{(\textrm{exp})}_{\mu}-a^{(\textrm{SM})}_{\mu}=(26\pm 9.4)\times10^{-10}$, suggested that the SM does not strictly hold in the low energy region when the difference exceeds the calculation uncertainties of the Hadronic process and measurement error. This difference has been extensively analyzed by various approaches, such as supersymmetry \cite{supersymmetric model},  lepton flavor violation \cite{nonconservation of lepton}, extra dimensions \cite{Graesser:1999yg}, etc. \cite{etc}. However, there is no conclusive explanation for this observed deviation. 

In this context, we attempt to estimate the order of the muon anomalous magnetic moment by using a braneworld model (see \cite{Rubakov:2001kp} for recent reviews). The general formalism using higher dimensional physics was constructed by R. Sundrum \cite{Sundrum:1998sj}. This formalism suggests that the SM particles are constrained to live on the world volume of a (3+1)-dimensional hypersurface or a ``$3$-brane," while only gravity freely propagates in bulk space-time.

The most important aspect of this theory is that the metric and the vierbein are replaced by the induced metric and the induced vierbein, respectively \cite{Sundrum:1998sj,Akama:1982jy}. Hence, higher dimensional gravity can be discussed apart from the usual Kaluza-Klein (KK) theory. 
In general, the KK modes require periodic limits such as the torus structure in extra dimensions. However, the braneworld scenarios need not have these limits because the configuration of extra dimensions is determined by gravity, the position of branes, the cosmological constant, etc. In this paper, we adopt the factorizable $6$-dimensional braneworld model of the ADD type (the name is derived from the paper by N. Arkani-Hamed, S. Dimopoulos, and G. Dvali \cite{Arkani-Hamed:1998rs}), which has a theoretical motivation that explains the gauge hierarchy problem, i.e., the reason for the scale of electroweak symmetry breaking being so much smaller than the scales of quantum gravity or grand unification. This model has extra compact spaces, and it finds a simple exact solution to the Einstein equation, including explicit brane sources. Applying Gauss's law to this model, we have
\begin{align}
M_{Pl}^2=M^{4}_{f}V_{(2)}\label{fundamental mass},
\end{align}
where $M_{Pl}\approx 10^{19}$ $[\textrm{GeV}]$, $M_{f}$, and $V_{(2)}$ denote the $4$-dimensional Planck mass, $6$-dimensional fundamental Planck mass, and volume of the two extra dimensions, respectively. This relation will be crucial in our discussion because it can give $M_{f}\approx 1$ [TeV] for ${V_{(2)}^{1/2}}\approx r=0.1$ [$\textrm{mm}$]\footnote{The conversion factor is $1[\textrm{GeV}^{-1}]=2\cdot10^{-13}[\textrm{mm}]$.}. This implies that gravity would be unified with other forces on a TeV scale. On the other hand, there exists an established higher dimensional model, the RS-model \cite{Randall:1999ee}, which could possibly resolve the hierarchy problem. However, we shall specifically concentrate on the $6$-dimensional ADD-type model \cite{Sundrum:1998ns,Carroll:2003db}. This involves the two $3$-branes (i.e., our world and another world) and the $U(1)$ gauge field in the bulk. The model can realize a mechanism that does not require any fine-tuning between the brane tension and bulk parameters; this implies that the brane tension can be freely changed. This type of model is referred to as a self-tuning model \cite{Carroll:2003db,Nilles:2003km}, which is one of the simplest models for exploring the braneworld phenomenology and the effects of extra dimensions. 
  
This paper is organized as follows. Section \ref{section 2} introduces some basic notations. Section \ref{section 3} comprises a brief explanation of the 6-dimensional model and the scaling property of $4$-dimensional physics, section \ref{section 4} focuses on the brane-stretching effect and the estimation of the order of muon (g-2), and section \ref{section 5} is the conclusion. 

\section{Setup}\label{section 2}
The effective theory has presented a picture of the low-energy dynamics of a 3-brane universe, i.e., the SM particle is confined to the braneworld-volume topology as ${\bf M}_{4}$. Further, only gravity is free to move in bulk space-time with $d>4$ dimensions, ${\bf M}_{4}\times {\bf S}^{d-4}$ topology, where ${\bf S}^{d}$ denotes the $d$-sphere. The coordinates of bulk space are denoted by $X^{M}$, where the ones on the brane are denoted by $x^{\mu}$ and the extra dimensions by $y^{m}$. The curved bulk coordinate indices, which run over all dimensions, are denoted by uppercase Roman letters beginning from the middle: $M,N \cdots =0,\cdots d-1$; the indices denoted by Greek letters run over the first four dimensions: $\mu,\nu,\cdots=0,1,2,3$; and the indices denoted by lowercase Roman letters run over the remaining $d-4$ dimensions: $m,n,\cdots=4,\cdots d-1$. The local Lorentz indices in the bulk are similarly denoted: the indices denoted by uppercase Roman letters run over all dimensions: $A,B \cdots =0,\cdots d-1$; Greek letters run over the first four dimensions: $\alpha,\beta\cdots=0,1,2,3$; the indices denoted by lowercase Roman letters run over the remaining $d-4$ dimensions: $a,b \cdots =4,\cdots d-1$  (see Table \ref{Summary of notations}).
\begin{ruledtabular}
\begin{table}[t]   
\begin{tabular}{ccc}
                    &Bulk space-time    &Braneworld volume    \\ \hline
Coordinates       &$X^{M}(M=0,1,\cdots,d-1)$   &$x^{\mu}(\mu=0,1,2,3)$  
\end{tabular}
\vspace{4mm}
\begin{tabular}{cccc}
       &$0,1,\cdots,d-1$   &$0,1,\cdots,3$&$4,\cdots,d-1$    \\ \hline
Curved indices      &$M,N,\cdots$  &$\mu,\nu,\cdots$ &$m,n,\cdots$   \\ \hline
Local Lorentz indices      &$A,B,\cdots$   &$\alpha,\beta\cdots$&$a,b,\cdots$  
\end{tabular}
\caption{Summary of notations}\label{Summary of notations}
\end{table}
\end{ruledtabular}

The bulk metric $G_{MN}$ describes the fundamental gravitational degrees of freedom. The Lorentz metric in the bulk is $\eta_{AB}$, and vielbein is $E^{A}_{M}(X)$. The bulk and Lorentz metrics are related by the following equation: 
\begin{align}
E^{A}_{M}(X)\eta_{AB}E^{B}_{N}(X)&=G_{MN}(X)\label{vielbein equation 1}\\
E^{A}_{M}(X)G^{MN}(X)E^{B}_{N}(X)&=\eta^{AB}.\label{vielbein relation 2}
\end{align}
The bulk coordinates occupied by a point $x$ on the brane are denoted by $Y^{M}(x)$. However, since the theory has reparametrization invariance, a different parametrization of the surface describing the brane $x\to x'(x)$ would lead to the same physics.
Therefore, it is necessary to identify the coordinates spanned by the brane with the first four bulk components in order to eliminate the non-physical components from $Y^{M}(x)$. Hence, we choose the gauge fixing condition 
\begin{align}
Y^{\mu}(x)=x^{\mu} .\label{gauge fix condition}
\end{align}
\section{Scaling property}\label{section 3}
We review the $6$-dimensional model with two brane sources in two extra dimensions, where the brane and the extra space have an ${\bf M}_{4}$ and an ${\bf S}^{2}$ topology, respectively \cite{Carroll:2003db,Aghababaie:2003wz,Nilles:2003km,Chen:2000at,Navarro:2003vw,Vinet:2004bk,Garriga:2004tq,Lee:2004vn,Mukohyama:2005yw}. The total action consists of the $6$-dimensional Einstein-Maxwell action and the two brane actions with negative tension. In this model, the stability of bulk geometry and brane fluctuations requires the negative tension brane. 

First, in order to obtain a background solution, we discuss the description that does not consider the localized fields on brane and brane fluctuations.
The effective action is shown to be as follows:
\begin{eqnarray}
S_{\textrm{total}}&=&S_{\textrm{branes}}+S_{6}\nonumber\\
&=&-T_{0}\int d^{4}x \sqrt{-g}-T_{1}\int d^{4}x \sqrt{-g}+\int d^{6}x \sqrt{-G}\left[M_{f}^{4}R_{6}-\Lambda_{6}-\frac{1}{4}F_{MN}^{2}\right]\label{total action}.
\end{eqnarray}
Here, $T_{i}$ $(i=0,1)$ denotes the brane tension, $M_{f}$ is the $6$-dimensional Planck mass, $\Lambda_{6}$ is the $6$-dimensional cosmological constant, and $F_{MN}$ is the $6$-dimensional $2$-form field strength. We can obtain the $6$-dimensional Einstein equation including brane sources by varying the action with respect to the $6$-dimensional metric. We consider that a brane is located on a conical singularity in the extra dimensions. Fortunately, in this scenario, we can easily obtain a solution that maintains a $4$-dimensional Minkowski space-time, because the equation can split into $4$- and $2$-dimensional components. Thus, we obtain the solution\footnote{The metric signature is diag $(+,-,-,-,-,-)$.}
 \begin{align}
ds^{2}_{6}=\eta_{\mu \nu}dx^{\mu}dx^{\nu}+\gamma_{mn}(y)dy^{m}dy^{n},\label{six dimensional metric}
\end{align}
and the solution for the equation of motion for $F_{MN}$ as
\begin{align}
F_{mn}=\sqrt{\gamma}B_{0}\epsilon_{mn}\label{magnetic flux},
\end{align}
where $B_{0}$ is a constant, $\gamma$ is the determinant of $\gamma_{mn}$, and $\epsilon_{mn}$ is a completely antisymmetric tensor, i.e., $\epsilon_{45}=-\epsilon_{54}=1$. Solution (\ref{magnetic flux}) denotes a magnetic flux through the compactified two extra dimensions.

In this background, the simplest technique to realize the stabilized bulk geometry would be to locate two fixed branes having identical tensions, $T_{0}=T_{1}$, at opposite poles of the spherical two extra dimensions. This condition can be ensured by imposing a ${\bf Z_{2}}$ symmetry at the equator \cite{Carroll:2003db}. Further, using the conformal symmetry, we can then obtain the solution 
\begin{align}
ds^{2}_{6}=\eta_{\mu \nu}dx^{\mu}dx^{\nu}+a_{0}^2(d\theta^{2}+\alpha^{2}\sin^{2}\theta d\phi^{2}),
\end{align}
if the parameters $B_{0}$ and $\lambda_{6}$ satisfy
\begin{align}
\frac{1}{a_{0}^{2}}=\frac{B_{0}^{2}}{2M^{4}_{f}}&,\quad\lambda_{6}=\frac{B^{2}_{0}}{2}\label{imposed condition}.
\end{align}
These relations are necessary to maintain a $4$-dimensional Minkowski space-time and spherical two extra dimensions.
The solution has the following relation on the conical singularity;
\begin{align}
\delta=2\pi(1-\alpha)=\frac{T_{0}}{2M_{f}^{4}},\label{deficit angle}
\end{align}
where $\delta$ is the deficit angle of the two extra dimensional sphere, and $\alpha$ is a dimensionless fixed parameter, $0<\alpha<1$. On the basis of a property in $2$-dimensional gravity \cite{Deser:tn}, the Einstein equation for the extra dimensional component presents a solution that removes a wedge from the sphere and was identified with opposite sides of the wedge.
Thus, the $4$-dimensional component remains exactly Lorentz invariant because the change in the tension affects only the geometry of the extra dimensions. This means that the tension can be freely changed since there is no fine tuning between bulk parameters and brane tension. This type of model is referred to as a self-tuning model.

In the following, we will briefly describe the manner in which this mechanism affects $4$-dimensional physics. The change in $T_{0}$ retains the regular part of the geometry and modifies only the singular part of the geometry, i.e., the deficit angle $\delta$ given by (\ref{deficit angle}). This results in a change in the bulk volume related to $M_{Pl}$ by (\ref{fundamental mass}). Hence, the change in $T_{0}$ signifies a change in $M_{Pl}$. Interestingly, a self-tuning model of this type can be constructed {\it only in six dimensions} \cite{Nilles:2003km}. 

Subsequently, we focus on fermion $\psi(x)$ and gauge field $A_{\mu}(x)$ and ignore scalar field on brane. However, prior to the discussing these behaviors, we should elaborate on a covariant derivative for the fermion. It behaves as a spin 1/2-spinor under the local Lorentz group. Lorentz generators of $n$-dimensional spinor representation are usually denoted as:
\begin{align}
\sigma_{(\alpha \beta)}=\frac{1}{4}\left[\gamma_{\alpha},\gamma_{\beta}\right],
\end{align}
where $\gamma_{\alpha}$ represents a set of Dirac matrices satisfying the following condition:
\begin{align}
\{\gamma_{\alpha},\gamma_{\beta}\}=2\eta_{\alpha\beta}.
\end{align}
The local Lorentz group on the brane is regarded as an internal $SO(3,1)$ group, which connects the Minkowski space with the curved space through the vielbein that satisfy Eqs. (\ref{vielbein equation 1}) and (\ref{vielbein relation 2}).
The covariant derivative that maintains the Lorentz and gauge symmetry for $\psi$ is 
\begin{eqnarray}
D_{\mu}&=&\partial_{\mu}-ieA_{\mu}-\frac{1}{2}\omega^{\alpha \beta}_{\mu}\sigma_{(\alpha \beta)},\\
\omega^{\alpha \beta}_{\mu}&=&\frac{1}{2}e^{\alpha \nu}(\partial_{\mu}e^{\beta}_{\nu}-\partial_{\nu}e^{\beta}_{\mu})+\frac{1}{4}e^{\alpha \nu}e^{\beta \sigma}(\partial_{\sigma}e^{\gamma}_{\nu}-\partial_{\nu}e^{\gamma}_{\sigma})e_{\gamma \mu}-(\alpha \leftrightarrow \beta)\label{spin connection}
\end{eqnarray}
\cite{Veltman}. Thus, the effective brane action is as follows:
\begin{eqnarray}
S_{\textrm{brane}}&=&\int d^{4}x \sqrt{-g}\left[-T_{0}+i\bar{\psi}e^{\mu}_{\alpha}\gamma^{\alpha}D_{\mu}\psi-m_{f}{\bar \psi}\psi-\frac{1}{4}g^{\mu \rho}g^{\nu \sigma}F_{\mu \nu}F_{\rho \sigma}+ \ \cdots \ \right],\label{3-brane lagrangian}
\end{eqnarray}
where the ellipsis represents the higher dimensional interactions that can be constructed with coefficients given by powers of 1/$M_{f}$, and $m_{f}$ is the mass parameter of the fermion in fundamental gravity. 

In the following, we discuss only the effect of the brane tension on $4$-dimensional physics (see \cite{Nilles:2003km} for details). The higher dimensional theory that results in a change in $M_{Pl}$ generates an effective theory depending on the change in $M_{Pl}$. Thus, the $4$-dimensional effective action consists of
\begin{align}
S_{\textrm{eff}}=M_{Pl}^{2}(T_{0})\int d^4 x\sqrt{-g}{R}_{4}+\int d^4 x\sqrt{-g}{{\cal L}}_{4}\label{effective gravity and brane action},
\end{align}
where $M_{Pl}$ is dependent on $T_{0}$ as follows:
\begin{align}
M_{Pl}^{2}(T_{0})=\biggl[1-\frac{T_{0}}{4\pi M^{4}_{f}}\biggr]M_{Pl}^{2}(0),
\end{align}
where $M_{pl}(0)$ represents the Planck mass in the absence of branes. When we rescale $\ g_{\mu \nu}={\tilde g}_{\mu \nu}/{\alpha}$, we obtain
\begin{eqnarray}
S_{\textrm{eff}}=M_{Pl}^{2}(0)\int d^4 x\sqrt{-{\tilde g}}{{\tilde R}}_{4}+\int d^4 x\sqrt{-\tilde{g}}{\tilde{{\cal L}}}_{4}(\alpha)\label{scaling property}.
\end{eqnarray}
It is obvious that the $\alpha$ dependence shifts from the Planck mass to the fields localized on the brane. Hence, after rescaling the fermions as $\ \psi=\alpha^{\frac{3}{4}}{\tilde \psi}$ on the basis of (\ref{3-brane lagrangian}), we obtain  
\begin{eqnarray}
{S}_{4}&=&\int d^4 x \sqrt{-\tilde{g}}\biggl[i{\bar{\tilde{\psi}}}e^{\mu}_{\alpha}\gamma^{\alpha}\left(\partial_{\mu}-ieA_{\mu}+\frac{1}{2}\omega^{\beta \gamma}_{\mu}\sigma_{(\beta \gamma)}\right){\tilde \psi}\nonumber\\
&&-\frac{m_{f}}{\sqrt{\alpha}}{\bar{\tilde \psi}}{\tilde \psi}-\frac{1}{4}g^{\mu \rho}g^{\nu \sigma}F_{\mu \nu}F_{\rho \sigma}\biggr].\label{3-brane lagrangian 3}
\end{eqnarray}
Based on the redefinition $\psi={\tilde \psi}$ and $g_{\mu\nu}=\tilde{g}_{\mu\nu}$, we recognize the action as invariant, except for the mass term. Thus, since $\alpha$ is the fixed parameter, we can regard $m_{f}/{\sqrt{\alpha}}$ as a physical mass $m$. As a result, the effect of bulk gravity does not become apparent in the $4$-dimensional world. This implies that if fermion is massless, the action becomes scale-invariant, i.e., the scale invariance is broken by fermion mass. The usual field theory also maintains this property. In the next section, we consider the effect of the brane tension and brane fluctuations. The $4$-dimensional field theory should be extended to a brane world that maintains this property. Further, we show that the scale transformation is instrumental in restricting the form of the induced metric. 

\section{Application to Muon (g-2)}\label{section 4}
\begin{figure}
\includegraphics[width=8cm,clip]{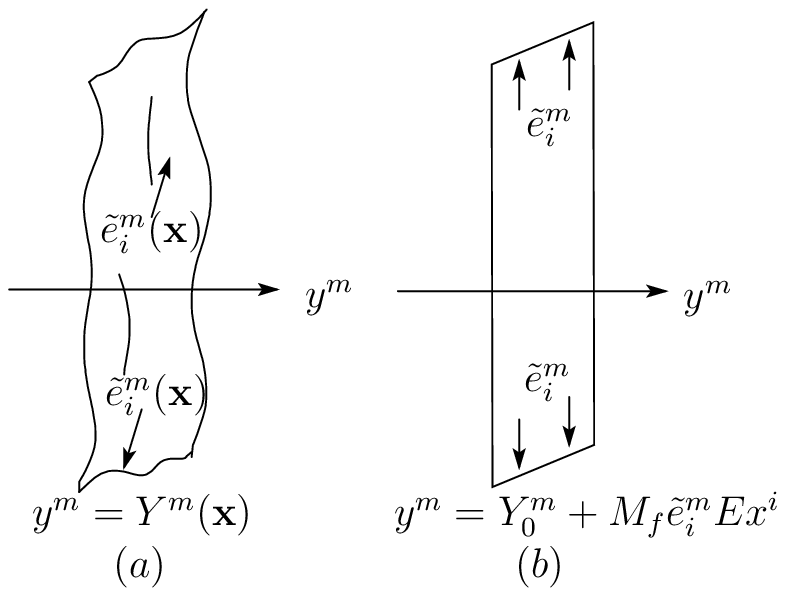}
\caption{The existence of brane separates bulk space-time into $4$D-space-time and extra space. (a) The static brane fluctuation generally allows the extra dimensional coordinate $y^{m}$ to acquire a dependence of spatial coordinates $x^{i}\ (i=1,2,3)$, $y^{m}=Y^{m}(\textbf{x})$. This implies the existence of a local frame given by the set of three basis vectors $\tilde{e}^{m}_{i}(\textbf{x})$ which are tangent to the spatial part of $3$-brane. (b) However, on a flat brane, we consider the basis vectors that do not depend on the local coordinate. Using these considerations, we can parametrize the coordinate $Y^{m}$ as $Y^{m}=Y^{m}_{0}+M_{f}{\tilde e}^{m}_{i}Ex^i$ based on the dimensional analysis; this is valid at the lower scale $E\ll  M_{f}$.}\label{figure1}
\end{figure}

We estimate the muon (g-2) deviation by assuming that brane fluctuations are static in time. The new compensation terms occur through the induced vierbein. This would lead to the possibility of compensating the magnetic moment which has a static property. Under gauge fixing condition (\ref{gauge fix condition}), the induced metric is as follows:
\begin{align}
g_{\mu \nu}=\eta_{\mu \nu}+\gamma_{mn}\partial_{\mu}Y^{m}\partial_{\nu}Y^{n}.\label{induced gravity}
\end{align} 
For simplicity, we suppose that off-diagonal components of the $6$-bein are zero, as shown below: 
\begin{align}
E^{A}_{M}(X)=\left(
  \begin{array}{cc}
     \delta_{\mu}^{\alpha}  &  0  \\
         0   & E^{a}_{m}   \\
  \end{array}
\right).
\end{align}
In order to obtain the induced vierbein on the brane, we use the following definition \cite{Sundrum:1998sj}:
\begin{align}
e^{\alpha}_{\mu}\equiv  R^{\alpha}_{A}E^{A}_{M}(X)\partial_{\mu}Y^{M}.
\end{align}
Thus, the induced vierbein obtains, up to the second order;
\begin{align}
e^{\alpha}_{\mu}=\delta^{\alpha}_{\mu}+\frac{1}{2}\gamma_{mn}\partial^{\alpha}Y^{m}\partial_{\mu}Y^{n}+O(\epsilon^4).\label{induced vierbein}
\end{align}
The expansion of $\sqrt{-g}$ of induced metric (\ref{induced gravity}) becomes
\begin{eqnarray}
\sqrt{-g}&=&1-\frac{1}{2}\partial^{\mu}Y^{m}\partial_{\mu}Y^{m}+\ \cdots.
\end{eqnarray}
The ellipsis consists of higher dimension terms of $\partial_{\mu}Y^{m}$ in pairs. When the above expansion is substituted into the minimal brane action
\begin{eqnarray}
S_{\textrm{brane}}&=&\int d^{4}x \sqrt{-g}\Bigl[-T_{0}+
{\cal L}_{\textrm{4}}(g_{\mu\nu})\Bigr]\label{3-brane action in gravity},
\end{eqnarray}
we obtain 
\begin{eqnarray}
S_{\textrm{brane}}&=&S_{\textrm{eff}}^{(0)}+S_{\textrm{eff}}^{(2)}+\cdots,\\
S_{\textrm{eff}}^{(0)}&=&\int d^{4}x \Bigl[-T_{0}+{\cal{L}}_{\textrm{4}}(\eta_{\mu\nu})\Bigr],\\
S_{\textrm{eff}}^{(2)}&=&\int d^{4}x \Bigl[\frac{T_{0}}{2}\partial_{\mu} Y^{m}\partial^{\mu}Y^{m}+\frac{1}{2}\partial_{\mu}Y^{m}\partial_{\nu}Y^{m}T^{\mu\nu}_{\textrm{4}}\Bigr],\label{3-brane lagrangian involed NG}
\end{eqnarray}
where $T^{\mu\nu}_{\textrm{4}}$ is the conserved energy-momentum tensor of matter fields evaluated in the $4$-dimensional Minkowski space-time. Considering the canonically normalized condition for $\partial_{\mu}Y^{m}$ in (\ref{3-brane lagrangian involed NG}),
we can put
\begin{align}
\partial_{\mu} Y^{m}\partial^{\mu} Y^{m}\rightarrow \frac{1}{T_{0}}\partial_{\mu} Y^{m}\partial^{\mu} Y^{m}.\label{canonically nomalization}
\end{align}
$Y^{m}$ is considered as the Nambu-Goldstone mode associated with the spontaneous isometry breaking due to the presence of the brane in bulk \cite{Sundrum:1998sj}\cite{Sundrum:1998ns}\cite{Hisano:1999bn,Bando:1999di,Kugo:1999mf,Dobado:2000gr}. 
Before discussing muon (g-$2$), the relation between the brane fluctuations and the scaling property mentioned in section \ref{section 3} should be noted. On the assumption that the change in $Y^{m}$ is static in time, the induced metric becomes
\begin{align}
g_{\mu \nu}=\left(
  \begin{array}{cc}
    1   & 0   \\
    0   & \eta_{ij}+\dfrac{1}{T_{0}}\gamma_{mn}\partial_{i}Y^m\partial_{j}Y^n  \\
  \end{array}
\right),\label{static induced metric}
\end{align}
where $i,j=1,2,3$:  indices are raised and lowered by the Euclidean metric $\delta_{ij}=-\eta_{ij}$. The $4$-dimensional field theory is scale-invariant for massless fermions and gauge fields, but not for massive fermions. We consider that the braneworld would preserve this property. Thus, induced metric (\ref{static induced metric}) requires the following rescaling for $\eta_{\mu\nu}\Rightarrow \eta_{\mu\nu}/\alpha$:
\begin{eqnarray}
g_{\mu\nu}&=&\left(
  \begin{array}{cc}
    1   & 0   \\
    0   & \eta_{ij}+\dfrac{1}{T_{0}}\gamma_{mn}\partial_{i}Y^m\partial_{j}Y^n  \nonumber\\
  \end{array}
\right)\\
&\Rightarrow& \frac{1}{\alpha}\left(
  \begin{array}{cc}
    1   & 0   \\
    0   & \eta_{ij}+\dfrac{1}{T_{0}}\gamma_{mn}\partial_{i}Y^m\partial_{j}Y^n  \\
  \end{array}
\right)\label{induced scaling}. 
\end{eqnarray}
However, the existence of the brane breaks the isometry symmetry. It denotes that the $6$-dimensional bulk is separated into the $4$-dimensional branes and $2$-dimensional extra dimensions. This implies that $\gamma_{mn}$ does not have the abovementioned transformation because we can rescale $G_{MN}\Rightarrow G_{MN}/\alpha$ if and only if no brane exists in the bulk. Therefore, in order to recover the scaling property, we restrict the form of $g_{ij}$:
\begin{align}
g_{ij}=\eta_{ij}+\eta_{ij}\dfrac{1}{T_{0}}H^2 M_{f}^2,\label{induced space}
\end{align}
where $H$ has a mass dimension of $1$. This form guarantees that the $\alpha$ dependence changes from $M_{Pl}(\alpha)$ into the fermion mass in the same way as action (\ref{3-brane lagrangian 3}). 
Subsequently, we present a solution $Y^{m}$ that satisfies (\ref{induced space}). The $Y^{m}$ equation of motion derived from effective action (\ref{3-brane lagrangian involed NG}) is written as
\begin{align}
\partial_{\mu}\Bigl[\partial^{\mu}Y^{m}+\frac{1}{T_{0}}\partial_{\nu}Y^{m}T^{\mu\nu}_{\textrm{4}}\Bigr]=0.\label{Y equation of motion}
\end{align}
When we introduce the dimensionless coordinate $Ex^{i}$ that characterizes the physical process at energy $E$, we parametrize $Y^{m}$ as follows (see Fig.\ref{figure1}):
\begin{align}
Y^{m}(\textbf{x})=Y^{m}_{0}+M_{f}{\tilde e}^{m}_{i}Ex^{i},\label{the solution of fluctuation}
\end{align}  
where $Y^{m}_{0}$ is a constant and the basis vectors
\begin{align}
\frac{\partial Y^{m}}{\partial x^{i}}=M_{f}E\tilde{e}^{m}_{i}
\end{align}
satisfy the completeness relation
\begin{align}
\gamma_{mn}\frac{\partial Y^{m}}{\partial x^{i}}\frac{\partial Y^{n}}{\partial x^{j}}=M_{f}^2E^2\eta_{ij},
\end{align} 
i.e.,
\begin{align}
\gamma_{mn}{\tilde e}^{m}_{i}{\tilde e}^{n}_{j}=\eta_{ij}.
\end{align}
The coordinate $Y^{m}$ (\ref{the solution of fluctuation}) satisfies (\ref{Y equation of motion}) and maintains (\ref{induced space}) as $H=E$. This implies that the spatial part of the brane is stretched due to brane fluctuations, whose magnitude depends on the energy scale of the physical process. This is physically plausible because under general relativity, space-time is not rigid but dynamical. 
In addition, solution (\ref{the solution of fluctuation}) is consistent with the general covariance of general relativity. Substituting (\ref{the solution of fluctuation}) into spin connection (\ref{spin connection}) via induced vierbein (\ref{induced vierbein}), we can directly obtain
\begin{eqnarray}
\omega^{\alpha\beta}_{\mu}&=&-\frac{1}{2}\partial_{\mu}\partial^{\beta}Y^{m}\partial^{\alpha}Y^{m}-\frac{1}{2}\partial_{\mu}\partial^{\alpha}Y^{m}\partial^{\beta}Y^{m}\nonumber\\
&=&0.
\end{eqnarray}
The vanishing of the spin connection denotes that the equation of motion for a fermion agrees with laws of special relativity. Therefore, solution (\ref{the solution of fluctuation}) supports Lorentz symmetry.

In the following, we will see that the brane-stretching effect generates the suitable order for muon (g-$2$). The variation of action (\ref{3-brane action in gravity}) with respect to $\bar{\psi}$ yields the equation of motion:
\begin{align}
\biggl[ie^{\mu}_{\alpha}\gamma^{\alpha}\left(\partial_{\mu}-ieA_{\mu}-\frac{1}{2}\omega^{\beta \gamma}_{\mu}\sigma_{(\beta \gamma)}\right)-m\biggr]\psi=0.\label{muon equation of motion}
\end{align}
Then, we perform a nonrelativistic approximation, i.e., the Schr\"odinger approximation. This is demonstrated in Appendix \ref{appendix1} by using the static induced metric (\ref{static induced metric}) and solution (\ref{the solution of fluctuation}).
Since the two extra dimensions give the relation
\begin{align}
M_{pl}^{2}=4\pi a_{0}^{2}\alpha M_{f}^{4}
\end{align}
by (\ref{fundamental mass}), we obtain the anomalous magnetic moment:
\begin{align}
a_{\mu}&=\frac{1}{T_{0}}E^2M_{f}^2\\
&=\frac{1}{4\pi M_{f}^{4}(1-\alpha)}E^2M_{f}^2\\
&=\frac{a_{0}^{2}\alpha}{M_{pl}^{2}(1-\alpha)}E^2M_{f}^2.\label{muon anomalous deficit angle}
\end{align} 
Finally, since we are interested in the physics at the muon scale ($E\approx 106[\textrm{MeV}]$) and $M_{f}\approx 1[\textrm{TeV}]$ for $a_{0}\approx0.1$ $[\textrm{mm}]$, we obtain the following:  
\begin{align}
a_{\mu}&\approx\frac{\alpha}{1-\alpha}\Big(\frac{0.1[\textrm{mm}]}{10^{19}[\textrm{GeV}]}\Big)^2\times\Big(106[\textrm{MeV}]\cdot 1[\textrm{TeV}]\Big)^2\\
&=\frac{\alpha}{1-\alpha}10^{-10}.
\end{align}
This result almost reproduces the deviation of the muon (g-2) measurement, except for the previous dimensionless factor. $\alpha$ may be determined by future studies on the self-tuning mechanism \cite{Aghababaie:2003wz,Nilles:2003km,Chen:2000at,Navarro:2003vw,Vinet:2004bk,Garriga:2004tq,Lee:2004vn,Mukohyama:2005yw}. However, it is important that we consider its behavior in the bound $0 < \alpha < 1$ because it is possible that $\alpha$ has an extreme value. If $\alpha \to 1$, muon (g-2) has a value greater than the experimental result. On the contrary, if $\alpha \to 0$, muon (g-2) has a small value. Moreover, it generates a large hierarchy between the fundamental parameter $M_{f}$ and $m_{f}$. Consequently, when $\alpha$ has a moderate value, this model would be capable of reproducing $\Delta a_{\mu}\equiv a^{(\textrm{exp})}_{\mu}-a^{(\textrm{SM})}_{\mu}=(26\pm 9.4)\times10^{-10}$.

As a side remark, from recent astrophysical research, it is known that the bounds on the mass of KK-gravitons \cite{Hall:1999mk} impose much tighter constraints on the radius of Large extra dimension. These suggest the exclusion of the TeV scale gravity. This indicates that we need to consider much more than the TeV scale. However, even in this case, if the order of $a_{0}$ is smaller than $0.1$ $[\textrm{mm}]$, the region $\alpha \to 1$ can give the appropriate (g-2) value if $\alpha$ is suitably selected. 

\medskip

\section{Conclusion}\label{section 5} 
This paper has presented a new approach according to which brane fluctuations compensate for the muon anomalous magnetic moment. The most important fact to be considered is that we have obtained a new potential term for the magnetic moment based on the assumption that brane fluctuations are static in time. This method reflects the effect of a novel classical contribution, namely, brane-stretching effect due to brane fluctuations, which is not based on the previously studied KK-gravitons \cite{Graesser:1999yg}. In particular, we would obtain a suitable order for $a_{\mu}$ in the $6$-dimensional model. This implies that the SM is consistently extended to the braneworld model that maintains the usual scaling property and Lorentz invariance for fermion. In future research, we should promote the investigation of $a_{\mu}$ by using the metric constructed by other higher dimensional models. Moreover, we can expect that the brane-stretching effect will evolve into different configurations in a very high energy. This may be related to the Lorentz violation \cite{Kostelecky:2000mm}. Since our study leaves a lot of issues to be discussed further, we are confident that this will be a crucial subject on which further research should be conducted. 

\begin{acknowledgments}
This study was partly supported by Iwanami F\={u}jyukai.
\end{acknowledgments}
\appendix*
\section{The sch{\"o}dinger approximation}\label{appendix1}
In this appendix, we demonstrate the non-relativistic approximation for fermion in the action (\ref{3-brane action in gravity}), and drive the magnetic moment. Varying the action with respect to $\bar{\psi}$, we obtain the equation of motion: 
\begin{align}
\biggl[ie^{\mu}_{\alpha}\gamma^{\alpha}\left(\partial_{\mu}-ieA_{\mu}-\frac{1}{2}\omega^{\beta \gamma}_{\mu}\sigma_{(\beta \gamma)}\right)-m\biggr]\psi=0, \label{muon equation of motion appendix}
\end{align}
where $e^{\mu}_{\alpha}$ is represented by (\ref{induced vierbein}). Operating on  
\begin{align}
\biggl[ie^{\mu}_{\alpha}\gamma^{\alpha}\left(\partial_{\mu}-ieA_{\mu}-\frac{1}{2}\omega^{\beta \gamma}_{\mu}\sigma_{(\beta \gamma)}\right)+m\biggr]
\end{align}
from the left, we get
\begin{widetext}
\begin{align}
&\biggl[g^{\mu \nu}\left(\partial_{\mu}
-ieA_{\mu}-\frac{1}{2}\omega^{({{\acute {\mathstrut \beta}}{\acute     {\mathstrut \gamma}}})}_{\mu}\sigma_{({{\acute {\mathstrut \beta}}{\acute{\mathstrut \gamma}}})}\right)\left(\partial_{\nu}
-ieA_{\nu}-\frac{1}{2}\omega^{({{{\mathstrut \beta}}{{\mathstrut    \gamma}}})}_{\nu}\sigma_{({{{\mathstrut \beta}}{{\mathstrut \gamma}}})}\right)+ie\sigma^{(\acute{\alpha}\alpha)}e^{\acute \mu}_{\acute \alpha}e^{\mu}_{\alpha}F_{\mu{\acute \mu}}\nonumber\\
&+\frac{1}{2}\sigma^{(\acute{\alpha}\alpha)}e^{\acute \mu}_{\acute \alpha}e^{\mu}_{\alpha}\Omega^{\beta \gamma}_{\mu{\acute \mu}}\sigma_{\beta \gamma}+\gamma^{\acute \alpha}\gamma^{\alpha}e^{\acute \mu}_{\acute \alpha}\partial_{{\acute \mu}}e^{\mu}_{\alpha}\left(\partial_{\mu}
-ieA_{\mu}-\frac{1}{2}\omega^{({{{\mathstrut \beta}}{{\mathstrut \gamma}}})}_{\mu}\sigma_{({{{\mathstrut \beta}}{{\mathstrut \gamma}}})}\right)
+m^{2}\biggr]\psi=0\label{Klein Gordon like}
\end{align}
\end{widetext}
by using the formula
\begin{align}
\gamma^{{\acute \alpha}}\gamma^{\alpha}=\eta^{{\acute \alpha}\alpha}+2\sigma^{({\acute \alpha}\alpha)},
\end{align}
and
\begin{widetext}
\begin{align}
&\sigma^{({\acute \alpha}\alpha)}e^{\acute \mu}_{\acute \alpha}e^{\mu}_{\alpha}\left(\partial_{\acute \mu}-ieA_{\acute \mu}-\frac{1}{2}\omega^{({{\acute {\mathstrut \beta}}{\acute{\mathstrut \gamma}}})}_{\mu}\sigma_{({{\acute {\mathstrut \beta}}{\acute{\mathstrut \gamma}}})}\right)\left(\partial_{\mu}-ieA_{\mu}-\frac{1}{2}\omega^{\beta \gamma}_{\mu}\sigma_{(\beta\gamma)}\right)\nonumber\\
&=\frac{1}{2}\sigma^{({\acute\alpha}\alpha)}e^{{\acute \mu}}_{{\acute \alpha}}e^{\mu}_{\alpha}\left(ieF_{\mu{\acute \mu}}+\frac{1}{2}\Omega^{\beta \gamma}_{\mu{\acute \mu}}\sigma_{\beta \gamma}\right)
\end{align}
\end{widetext}
where $F_{\mu \nu}\equiv\partial_{[\mu}A_{\nu]}$ and $\Omega^{\beta \gamma}_{\mu \nu}\equiv\partial_{[\mu}\omega^{\beta \gamma}_{\nu]}$.
Further, given the assumption that the change in $Y^m (x)$ is static in time, we obtain the induced metric 
\begin{align}
g_{\mu \nu}=\left(
  \begin{array}{cc}
    1   & 0   \\
    0   & \eta_{ij}+\gamma_{mn}\partial_{i}Y^m\partial_{j}Y^n  \\
  \end{array}
\right).
\end{align}
Thus, rewriting
\begin{align}
p^{\mu}=i\partial^{\mu}\quad,\quad A_{\mu}=(\phi, \vec{A}),
\end{align}
the (\ref{Klein Gordon like}) transforms into 
\begin{widetext}
\begin{eqnarray}
-\left(iE+ie\phi-\frac{1}{2}\omega^{{{{\mathstrut \beta}}{{\mathstrut \gamma}}}}_{0}\sigma_{({{{\mathstrut \beta}}{{\mathstrut \gamma}}})}\right)^2\psi&=&\biggl[g^{ij}\left(-ip_{i}
-ieA_{i}-\frac{1}{2}\omega^{({{\acute {\mathstrut \beta}}{\acute {\mathstrut \gamma}}})}_{i}\sigma_{({{\acute {\mathstrut \beta}}{\acute{\mathstrut \gamma}}})}\right)\left(-ip_{j}
-ieA_{j}-\frac{1}{2}\omega^{({{{\mathstrut \beta}}{{\mathstrut    \gamma}}})}_{j}\sigma_{({{{\mathstrut \beta}}{{\mathstrut \gamma}}})}\right)\nonumber
\\&&-ie\sigma^{ij}e^{k}_{i}e^{l}_{j}F_{kl}-\frac{1}{2}\sigma^{ij}e^{k}_{i}e^{l}_{j}\Omega^{\beta \gamma}_{kl}\sigma_{\beta\gamma}\nonumber\\&&+\gamma^{i}\gamma^{j}e^{k}_{i}\partial_{k}e^{l}_{j}\left(-ip_{l}-ieA_{l}-\frac{1}{2}\omega^{(\beta \gamma)}_{i}\sigma_{\beta\gamma}\right)+m^{2}\biggr]\psi\label{eigenvalue equation}
\end{eqnarray}
\end{widetext}
where $i,j,k,l=1,2,3$ and $E$ represents the energy eigenvalue. Putting $E=m+W$ where $m$ is the rest energy, the L.H.S of (\ref{eigenvalue equation}) is as follows:
\begin{eqnarray}
{\textrm L}.{\textrm H}.{\textrm S}&=&\biggl[m^2+2m\left(W+e\phi+\frac{i}{2}\omega^{{{{\mathstrut \beta}}{{\mathstrut \gamma}}}}_{0}\sigma_{({{{\mathstrut \beta}}{{\mathstrut \gamma}}})}\right)\nonumber\\
&&+\left(W+e\phi+\frac{i}{2}\omega^{{{{\mathstrut \beta}}{{\mathstrut \gamma}}}}_{0}\sigma_{({{{\mathstrut \beta}}{{\mathstrut \gamma}}})}\right)^{2}\biggr]\psi\label{eigenvalue equation 2}.
\end{eqnarray}
In addition, we assume that $W \ll m$, i.e., the energy due to a magnetic field is extremely small. In this case, dividing both the L.H.S and R.H.S of (\ref{eigenvalue equation}) by $2m$ so as to ignore the last term in (\ref{eigenvalue equation 2}), we obtain 
\begin{widetext}
\begin{eqnarray}
W\psi&=&\frac{1}{2 m}\biggl[g^{ij}\left(-ip_{i}-ieA_{i}-\frac{1}{2}\omega^{({{{\acute{\mathstrut \beta}}{\acute{\mathstrut \gamma}}}})}_{i}\sigma_{(\beta\gamma)}\right)\left(-ip_{j}-ieA_{j}-\frac{1}{2}\omega^{\beta\gamma}_{j}\sigma_{(\beta\gamma)}\right)\nonumber\\
&&-ie\sigma^{ij}e^{k}_{i}e^{l}_{j}F_{kl}-\frac{1}{2}\sigma^{ij}e^{k}_{i}e^{l}_{j}\Omega^{\beta\gamma}_{kl}\sigma_{\beta\gamma}+\gamma^{i}\gamma^{j}e^{k}_{i}\partial_{k}e^{l}_{j}\left(-ip_{l}-ieA_{l}+\frac{1}{2}\omega^{\beta\gamma}_{j}\sigma_{(\beta\gamma)}\right)\biggr]\psi\nonumber\\&&
-\left(e\phi+\frac{i}{2}\omega^{\beta\gamma}_{0}\sigma_{(\beta\gamma)}\right)\psi.
\label{eigen value equation3}
\end{eqnarray}
\end{widetext}
This is the eigenvalue equation for a charged particle in a magnetic field and gravity. From this equation, we can ascertain the energy shift term, which is produced by the following interaction: 
\begin{align}
\frac{\partial W}{\partial H_{i}}H_{i}=\frac{-ie}{2m}\sigma^{ij}e^{k}_{i}e^{l}_{j}F_{kl}.\label{muon magnetic moment}
\end{align}
Hence, when evaluating Eq. (\ref{muon magnetic moment}) by using $e^{k}_{i}$ which is the inverse of Eq. (\ref{induced vierbein}) and the solution (\ref{the solution of fluctuation}), we obtain 
\begin{align}
\frac{\partial W}{\partial H_{i}}H_{i}=\frac{-e}{2m}\biggl[\Bigl(2+2\frac{E^2M_{f}^2}{T_{0}}\Bigr)\frac{{\vec \sigma}}{2}\cdot{\vec H}\biggr]\label{anomalous magnetic moment appendix}
\end{align}
where $F_{23}=-F_{32}=H_{1}$, $F_{31}=-F_{13}=H_{2}$, and $F_{12}=-F_{21}=H_{3}$. The parenthesis of the term proportional to ${\vec \sigma}/2\cdot{\vec H}$ represents the magnetic moment.

\end{document}